\documentclass{nature}

\usepackage{graphicx}
\title{Cellular network entropy as the energy potential in Waddington's differentiation landscape}

\author{Christopher R. S. Banerji$^{1,2}$, Diego Miranda-Saavedra$^3$ , Simone Severini$^{2,4}$, Martin Widschwendter$^5$, Tariq Enver$^6$, Joseph X. Zhou$^7$ \& Andrew E. Teschendorff$^{1,2,8,*}$
}

\begin{document}
\maketitle

\begin{affiliations}
 \item Statistical Cancer Genomics, Paul O'Gorman Building, UCL Cancer Institute, University College London, 72 Huntley Street, London WC1E 6BT, United Kingdom.
 \item Centre for Mathematics and Physics in the Life Sciences and Experimental Biology, University College London, London WC1E 6BT United Kingdom.
 \item Bioinformatics and Genomics Laboratory, World Premier International (WPI) Immunology Frontier Research Center (IFReC), Osaka University, Osaka, Japan.
 \item Department of Computer Science, University College London, Gower Street, London WC1E 6BT, United Kingdom.
 \item Department of Women's Cancer, University College London, London WC1E 6BT, United Kingdom.
 \item UCL Cancer Institute, University College London, 72 Huntley Street, London WC1E 6BT, United Kingdom.
 \item Institute for Systems Biology, 401 Terry Avenue North, Seattle, WA 98109-5234, USA.
 \item CAS-MPG Partner Institute for Computational Biology, Chinese Academy of Sciences, Shanghai Institute for Biological Sciences, 320 Yue Yang Road, Shanghai 200031, China.
\end{affiliations}

\begin{addendum}
 \item[$^*$] Corresponding author: a.teschendorff@ucl.ac.uk
\end{addendum}

\begin{abstract} 
Differentiation is a key cellular process in normal tissue development that is significantly altered in cancer. Although molecular signatures characterising pluripotency and multipotency exist, there is, as yet, no single quantitative mark of a cellular sample's position in the global differentiation hierarchy. Here we adopt a systems view and consider the sample's network entropy, a measure of signaling pathway promiscuity, computable from a sample's genome-wide expression profile. We demonstrate that network entropy provides a quantitative, in-silico, readout of the average undifferentiated state of the profiled cells, recapitulating the known hierarchy of pluripotent, multipotent and differentiated cell types. Network entropy further exhibits dynamic changes in time course differentiation data, and in line with a sample's differentiation stage. In disease, network entropy predicts a higher level of cellular plasticity in cancer stem cell populations compared to ordinary cancer cells. Importantly, network entropy also allows identification of key differentiation pathways. Our results are consistent with the view that pluripotency is a statistical property defined at the cellular population level, correlating with intra-sample heterogeneity, and driven by the degree of signaling promiscuity in cells. In summary, network entropy provides a quantitative measure of a cell's undifferentiated state, defining its elevation in Waddington's landscape.
\end{abstract}

\section*{Introduction} 
The observed diversity of mature cells and human tissues arises as a result of a complex, intricate program of cellular differentiation, ultimately originating from (pluripotent) embryonic stem cells \cite{Keller2005}. Although systems biology principles underpinning the transitions between specific cellular states, such as pluripotency and progenitor states, are in the process of being elucidated \cite{Zhou2011,Zhou2011tg,Ladewig2013,Heinaniemi2013}, much remains to be learned. In the case of hematopoiesis, one of the best understood developmental systems, the full repertoire of transcription factors and signaling pathways dictating cell-fate is still unknown \cite{Chang2008,Pina2012,Kohn2012,Rodrigues2012,Heinaniemi2013}. Other studies have focused on characterising the pluripotent and progenitor states in terms of genome-wide gene expression \cite{Chen2008,Mueller2008,Mikkelsen2008,Wong2008,Mueller2011,Palmer2012}, DNA methylation and chromatin state profiles \cite{Meissner2008,Ernst2011,Zhu2013,Smith2013,Bock2012}. Although these molecular signatures can discriminate cells of specific differentiation stages from each other, there is, as yet, no single quantitative measure that can correctly place a sample within the global differentiation hierarchy. Rephrased in the context of Waddington's differentiation landscape \cite{Waddington1957}, we do not yet have a molecular measure that can represent the energy potential, i.e. the elevation, in Waddington's landscape.\\
Recently, it has been proposed that pluripotency, and more generally, the undifferentiated state, is an emergent statistical property of a population of cells \cite{MacArthur2013,Furusawa2009,Furusawa2012}, not well-defined at the single-cell level. Specifically, it has been argued that high cellular diversity underpins the pluripotent or multipotent capacity of stem cell populations, with differentiated cell populations representing a more uniform synchronised state \cite{MacArthur2013}. Motivated by this, we here explore a system's property of a cellular sample, called {\it network entropy}, in the context of cellular differentiation. At the single-cell level network entropy can be thought of as an approximate measure of signaling pathway promiscuity \cite{Teschendorff2010bmc,West2012,MacArthur2013,Li2013biophy}. Thus, a highly undifferentiated cell, such as a pluripotent stem cell, would have a high network entropy since it must maintain the option to initiate the activation of a wide number of different signaling pathways associated with commitment to diverse cell fates \cite{Chang2008}. In contrast, a terminally differentiated cell would have a low network entropy, since it must maintain activation of a few pathways specific to their fate. At the population level, high network entropy would thus imply increased cellular heterogeneity, since the increased signaling promiscuity results in an increased stochasticity across single cells. Thus, we posited that network entropy would provide a direct molecular correlate of the undifferentiated state of a cellular sample, allowing us to place an arbitrary sample at its appropriate elevation in Waddington's landscape.\\
To test our hypothesis, we here compute sample-specific network entropies for a large number of gene expression data sets relevant to cellular differentiation, reprogramming and cancer, encompassing over 800 samples, including cell-lines and primary tissue. Our main key findings are: (i) network entropy is a highly accurate discriminator of pluripotent and non-pluripotent cell-types, (ii) it can further discriminate cellular states of varying degrees of multipotency within distinct lineages, (iii) it provides a more robust and general measure of a cell's position in the global differentiation hierarchy than gene expression signatures, and does so independently of cell proliferation, and (iv) it predicts a higher cellular heterogeneity in cancer stem cells compared to ordinary cancer cells.

\section*{Results}

\subsection{Construction and rationale of network entropy}

To compute network entropy requires estimation of the signaling/interaction probabilities of proteins in a given sample. Thus, we integrated the gene expression profile of a given sample with a comprehensive protein interaction signaling network (PIN) ({\bf see SI },\cite{Cerami2011}), using the mass-action principle to construct a sample-specific stochastic matrix $p_{ij}$ where $i$ and $j$ label two distinct genes. The stochastic matrix provides a rough proxy for the interaction probabilities present in the given sample and its construction is based on the assumption that two genes known to interact at the protein level will have a greater interaction probability when they are both highly expressed ({\bf see SI }). From the stochastic matrix, the network entropy can be calculated as the entropy rate \cite{Latora1999,GomezGardenes2008}
\begin{equation}
S_R=\sum_i{\pi_iS_i}
\end{equation}
where $S_i$ is the local entropy of node (gene/protein) $i$ and where $\pi_i$ is the $i$'th element of the stationary distribution of $p_{ij}$ (i.e. $\pi p=\pi$, see {\bf Methods, SI }). Thus, the entropy rate gives a steady-state average measure of the uncertainty (or promiscuity) in signaling information flow over the network. To facilitate comparison of entropy rates obtained from samples profiled on different expression arrays, values were always normalised to the maximum possible entropy rate of a given integrated network ({\bf SI , fig.S1}).\\
We posited that the entropy rate of a sample (e.g. a cell-line), as computed above, would capture the average level of signaling pathway promiscuity and hence of the cellular heterogeneity in the sample. Under this model, highly undifferentiated and plastic cells, such as stem cells, would be characterised by a state of high network entropy, allowing them the option to differentiate into diverse cell lineages ({\bf Fig.1A}). Similarly, since differentiation implicates activation of specific molecular signaling pathways, this activation would lead to a reduction in the uncertainty/promiscuity of information flow, i.e. a low entropy state ({\bf Fig.1A}).

\begin{figure}[th]
\begin{center}
\includegraphics[scale=0.75]{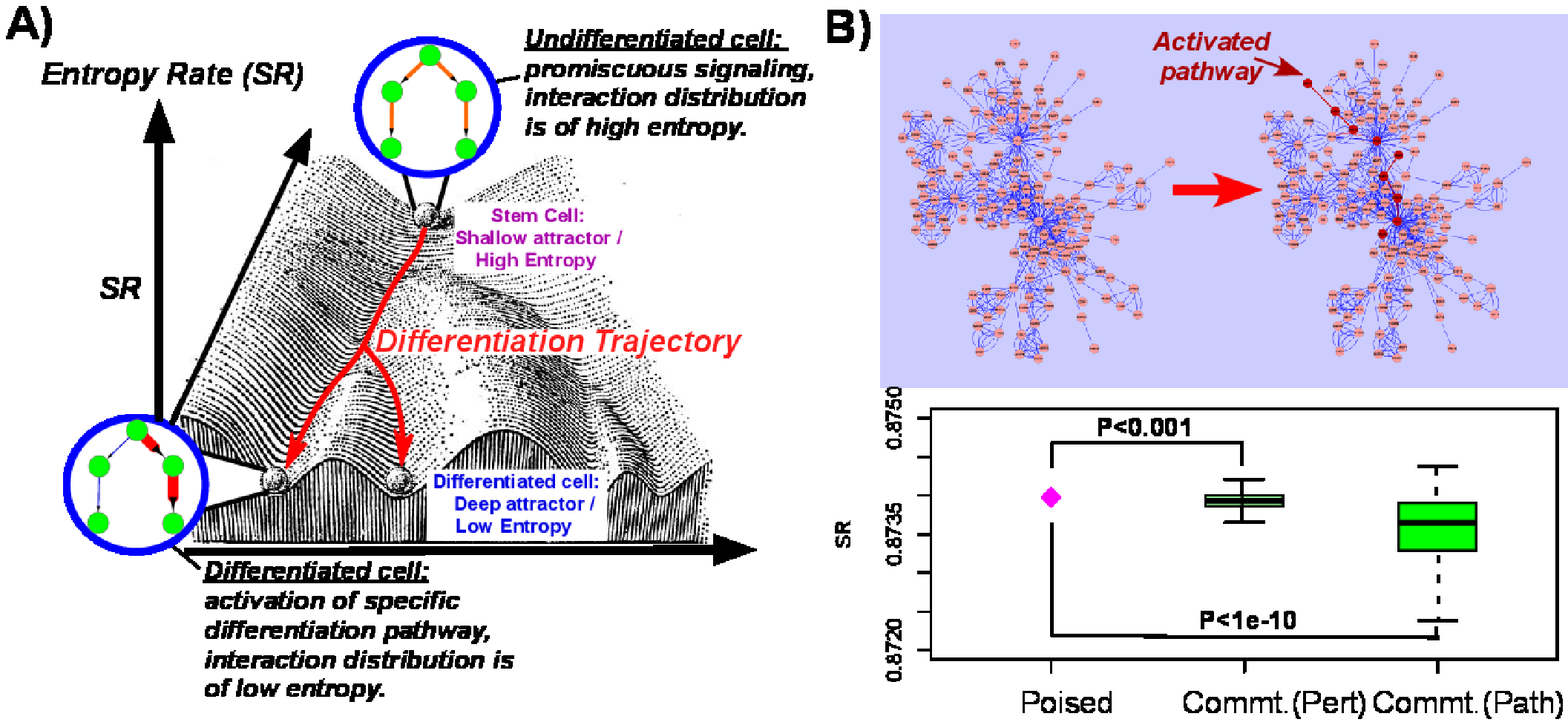}
%
%
\caption{{\bf Network entropy as the energy potential in Waddington's landscape: A)} Illustration of network entropy's role in cellular differentiation. The z-axis represents the network entropy rate (SR) of a cell, which is a measure of the promiscuity/redundancy in the signaling patterns within the cell. The two-dimensional plane spanned by the x-and-y axis represents gene expression state/phase space. {\bf B)} Simulation of pathway activation in a realistic protein interaction network (only a small subnetwork is shown). In the left, edge weights are defined equally, so that the random walk on the network is unbiased. On the right, a specific pathway is activated by increasing the relative weights of edges connecting the genes in the pathway (shown in dark red). Lower panel compares the entropy rate (SR) of the unbiased state, representing a highly promiscuous poised cellular state (magenta diamond), to the entropy rates obtained by separately activating each individual gene in the network ($>1000$ perturbations, ``Commt(Pert.)''), and to the entropy rates obtained by activating whole signal transduction pathways (100 pathways, ``Commt.(Path)''). Binomial test P-values are given.}
\label{fig:1}       
\end{center}
\end{figure}

As a proof of concept that the entropy rate does indeed measure the level of signaling promiscuity we first devised a simulation model ({\bf SI }). We compared the entropy rate of our PIN with weights defined by a uniform stochastic matrix (i.e. one with $p_{ij}\propto 1/k_i$ where $k_i$ is the degree of node $i$) representing a promiscuous poised state, to the entropy rate obtained by randomly activating individual genes and specific signal transduction pathways in the network ({\bf SI , Fig.1B}). In the case where individual genes were activated, this led, in approximately 70$\%$ of perturbations, to a reduction in the global entropy rate (Binomial $P<0.001$, {\bf Fig.1B}). However, in the case where whole signaling pathways were activated, the reduction in the entropy rate was observed in $85\%$ of cases (Binomial $P<10^{-10}$, {\bf Fig.1B}), consistent with a substantially lower uncertainty in the information flow.

\subsection{Network entropy quantifies the level of multipotency}
Based on the simulation results, we sought to determine if network entropy could discriminate biological samples that differ in terms of their signaling promiscuity. Thus, we computed the network entropy rate of samples in the ``stem cell matrix'' (SCM), a compendium of over 219 samples (mostly cell-lines), all profiled with the same Illumina arrays, 59 of which were deemed pluripotent, with the rest (160) deemed non-pluripotent \cite{Mueller2008}. We observed that network entropy was significantly higher in the cell-lines deemed pluripotent ($P<10^{-10}$, {\bf Fig.2A}). To provide an independent benchmark we also computed a t-test based pluripotency score (TPSC, {\bf SI }), constructed from an independent 19-gene pluripotency expression signature, containing important pluripotency markers such as {\it NANOG} and {\it LIN28A} \cite{Mikkelsen2008}. The TPSC pluripotency score was also significantly higher in the pluripotent cell lines ({\bf SI , fig.S2}), and both measures were significantly correlated, confirming that network entropy is indeed a marker of pluripotency ({\bf Fig.2B}). In an independently generated data set profiling 107 human embryonic and 52 induced pluripotent stem cell lines, as well as 32 differentiated tissue samples \cite{Nazor2012}, the entropy rate achieved 100$\%$ accuracy in discriminating pluripotent from differentiated samples ({\bf Figs.2C-D}). Crucially, all these results were independent of cell proliferation, as we verified by removing cell proliferation and cycling genes \cite{BenPorath2008} from the network ({\bf see SI , figs.S3-S4}). Furthermore, passage number and sex did not have noticeable effects on the entropy rate as assessed in 107 human embyronic stem cell (hESC) lines ({\bf SI , fig.S5}). Consistent with network entropy being a marker of pluripotency we observed that induced pluripotent stem cell samples exhibited high entropy values, similar to that of hESCs, and significantly higher than that of their parental differentiated cells ($P<0.0001$, {\bf SI , figs.S6-S7}).

\begin{figure}[ht]
\begin{center}
\includegraphics[scale=0.9]{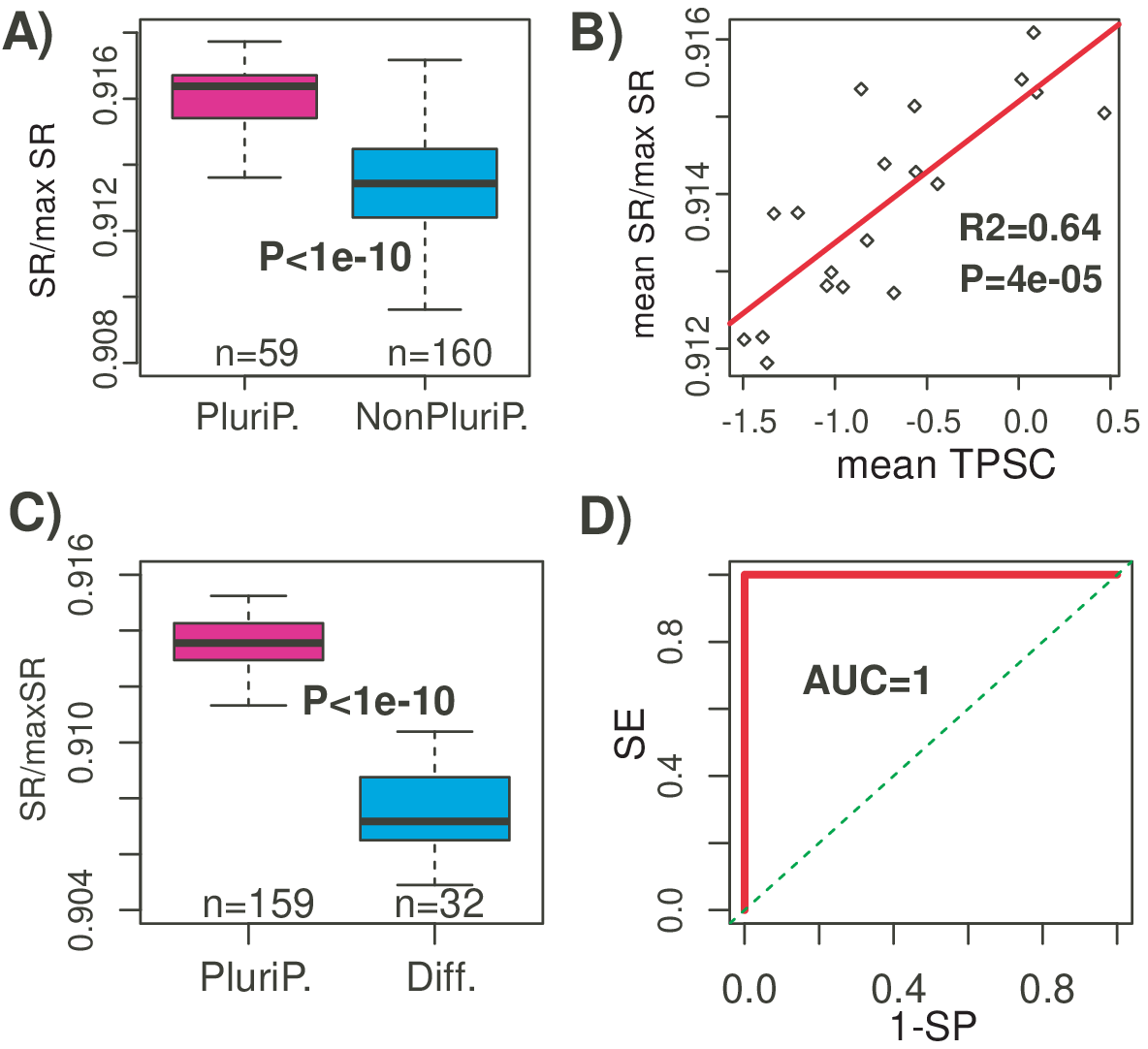}
%
%
\caption{{\bf Network entropy correlates with pluripotency:} {\bf A)} Normalised entropy rates (SR/maxSR, y-axis) between the 59 pluripotent and 160 non-pluripotent cell-lines from the SCM compendium (219 samples). P-value is from a Wilcoxon rank sum test. {\bf B)} Scatterplot of the entropy rate vs pluripotency score, where values for replicates of each cell type have been averaged. Linear regression P-value and $R^2$ value are given. {\bf C)} Normalised entropy rates (SR/maxSR, y-axis) between the 159 pluripotent and 32 differentiated samples from the SCM2. P-value is from a Wilcoxon rank sum test. {\bf D)} Corresponding ROC curve plus AUC of network entropy discriminating pluripotent from differentiated cells.}
\label{fig:2}       
\end{center}
\end{figure}

Next, we compared the network entropy of hESCs to that of committed but multipotent cell types, including neural stem cells (NSCs), hematopoietic stem cells (HSCs) and mesenchymal stem cells (MSCs). Confirming our hypothesis, all of these stem cell types exhibited entropies which were lower than that of hESCs/iPSCs, but higher than their differentiated progeny ({\bf Fig.3A}, {\bf SI , S8-S9}). Thus, network entropy can discriminate cells within a lineage according to their differentiation status. To test this further, in a combined haematological data set \cite{Watkins2009}, encompassing a number of different blood cell types including differentiated types (e.g. monocytes), and less differentiated ones (e.g. CD34+ HSCs and erythroblasts/megakaryocytes), network entropy recapitulated a differentiation hierarchy consistent with prior knowledge \cite{Goldfarb2001,Miranda-Saavedra2008} (see {\bf SI , fig.S10}). Importantly, we observed that network entropy was a relatively robust measure, being fairly insensitive to the normalisation or platform used ({\bf SI , figs.S8-S11}), although in the case of MSCs biological variations were evident ({\bf SI , figs.S8}) \cite{Zipori2006,Krinner2010}.

 \subsection{Network entropy is reduced during differentiation}
If network entropy is a general measure of the undifferentiated state of cells, it ought to exhibit dynamic changes in time course differentiation data. To this end, we considered expression data of differentiated retinal pigment epithelial cells, which were induced to de-differentiate, followed by a period of re-differentiation ({\bf SI }). Remarkably, network entropy increased upon de-differentiation, reaching a maximum, with values subsequently dropping upon re-differentiation ({\bf Fig.3B}). As another example, we considered a time course data set consisting of human promyelocytic leukemia progenitor (HL60) cells, differentiating into neutrophils \cite{Huang2005}. There were two separate time courses, using distinct stimuli to induce differentiation of HL60 cells. In both cases, network entropy was significantly reduced with time (ATRA stimulus, $R^2=0.96$, $P<10^{-8}$, {\bf Fig.3B}). Once again, these dynamic changes were independent of cell-proliferation ({\bf SI , fig.S12}). 

\begin{figure}[ht]
\begin{center}
\includegraphics[scale=0.75]{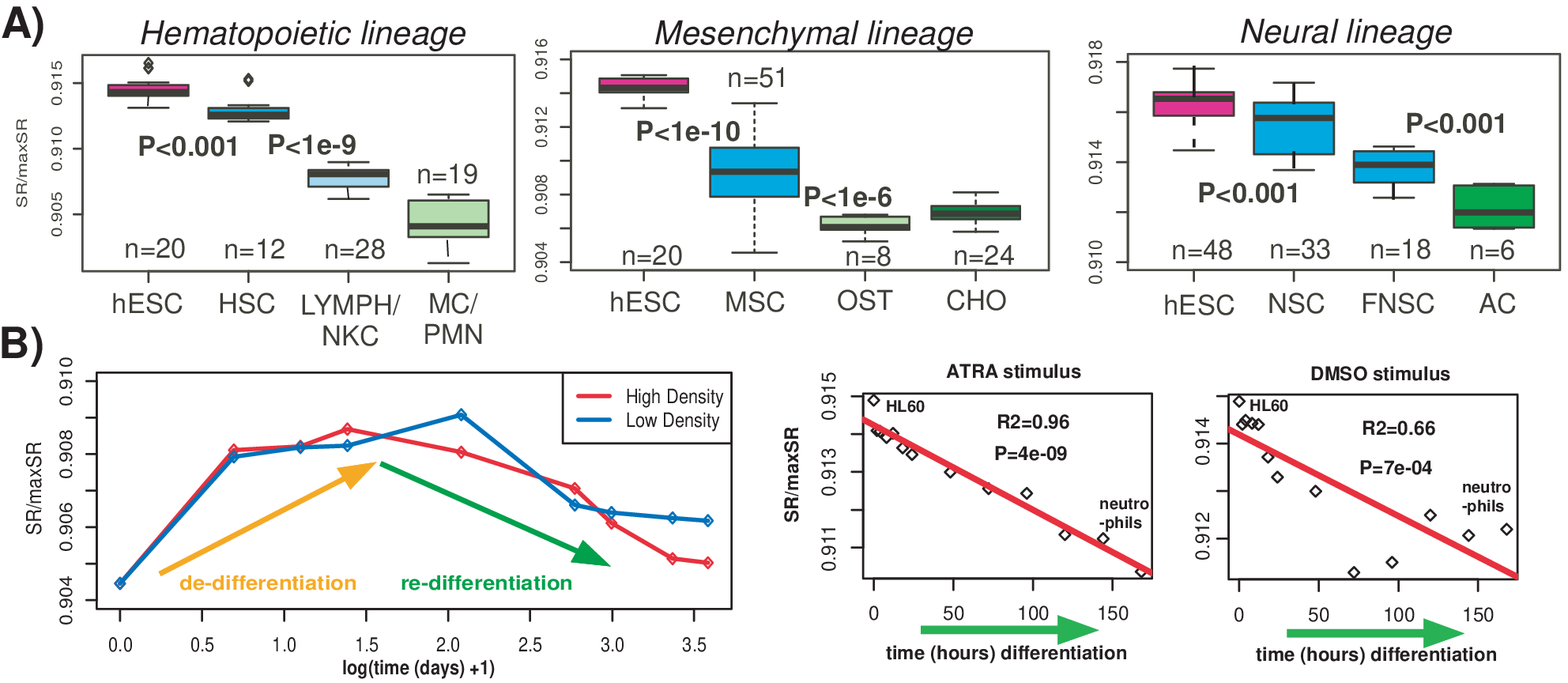}
%
%
\caption{{\bf Network entropy marks differentiation potential:} {\bf A) Multi-lineage analysis:} Left panel: Comparison of normalised network entropy values of hESCs, hematopoietic stem cells (HSCs), T \& B-cell lymphocytes plus natural killer cells (LYMPH/NKC), and monocytes plus neutrophils (MC/PMN). Middle panel: Comparison of normalised network entropy values of hESCs, mesenchyma stem cells (MSCs) and differentiated osteoblasts (OST) and chondrocytes (CHO). Right panel: Comparison of normalised network entropy values of hESCs, neural stem cells (NSCs) derived from the hESCs, fetal neural stem cells (FNSC) and primary astrocytes (AC), as derived from the SCM compendium (Illumina arrays). Wilcoxon rank sum test P-values between consecutive groups in the differentiation hierarchy are given. {\bf B) Dynamic changes in network entropy:} Left panel: Network entropy changes in a time course de-differentiation and re-differentiation experiment of retinal pigment epithelium (RPE), with cell density indicating the initial plating density of RPE cells. Right panels: Network entropy rate (SR/maxSR, y-axis) changes of HL60 leukemic progenitor cells against time from initial stimulus with either ATRA or DMSO. The data points on the left indicate the less differentiated HL60 cells, whereas the ones on the far right represent differentiated neutrophils. We provide the $R^2$ values and associated P-values from a linear regression. 
}
\label{fig:3}       
\end{center}
\end{figure}

\subsection{Network entropy discriminates cancer stem cells, cancer and normal cells}

Differentiation is a key distinctive feature of cancer and normal cells, with cancer representing a less differentiated and more heterogeneous state. Confirming this, network entropy was consistently higher in cancer tissue compared to normal cells, across four different tissue types, with cancer cell-lines exhibiting  even higher values ({\bf Fig.4A}). We further analysed an expression data set profiling putative cancer stem cells (CSCs) and their parental cancers across a number of different tissues \cite{Yu2012}. This showed that CSCs exhibited a marginally higher network entropy than their non-stem like counterparts, consistent with the view that CSCs retain a higher level of plasticity ({\bf Fig.4B}).\\
Interestingly, comparing the network entropy of hESCs to teratocarcinomas and germ cell tumours (all from the SCM and all deemed pluripotent), revealed marginally higher values in the hESCs ({\bf SI , fig.S13}). This pattern of higher network entropy in normal stem cells was also seen in the non pluripotent context: for instance, the network entropy of HSCs and NSCs was, in general, higher than that of leukemic stem and glioma stem cells, respectively ({\bf SI , fig.S13-S14}). Thus, while CSCs and ordinary cancer cells exhibit significantly increased cellular heterogeneity compared to normal differentiated tissue, CSCs do not appear to exhibit higher values relative to their normal stem cell counterparts, and even appear to show reduced levels of entropy compared to normal stem cells.

\begin{figure}[ht]
\begin{center}
\includegraphics[scale=0.75]{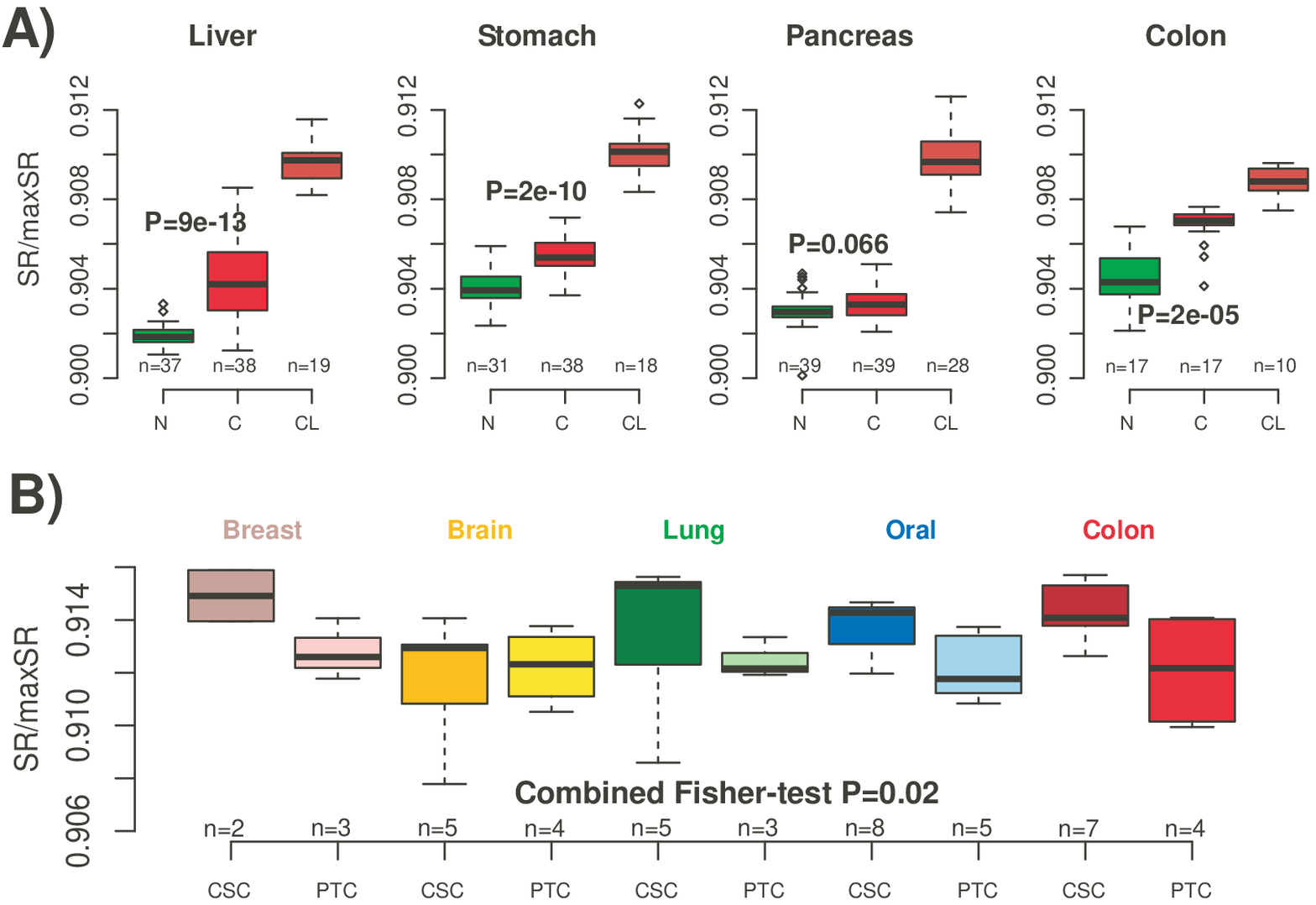}
%
%
\caption{{\bf Network entropy discriminates cancer stem cells, cancer and normal tissue:} {\bf A)} Comparison of normalised entropy rates (SR/maxSR) between normal and cancer tissue, as well as cancer cell lines, across four different tissue types, as indicated. {\bf B)} Comparison of normalised entropy rates between putative cancer stem cells (CSC) and their parental tumour cell lines (PTC) for five different tissue types. Combined Fisher t-test P-value is given.}
\label{fig:4}       
\end{center}
\end{figure}

\subsection{Dynamic changes in local network entropy identifies key differentiation genes and pathways}
To demonstrate that the dynamic changes in entropy can be related to changes in activation of specific pathways, we considered, as a proof of principle, the case of Notch-signaling. Notch signaling is inactive yet inducible in the pluripotent state, with activation normally associated with differentiation \cite{Noggle2006,Yu2008,Meier-Stiegen2010,Liu2010,Bigas2012,Blank2008}. Thus, essential components of the Notch signaling pathway should exhibit a lower network entropy in the non-pluripotent state. Using data from the stem cell matrix \cite{Mueller2008}, we were able to confirm this for 12 of the 13 Notch pathway genes ({\bf SI , figs.S15-S16}). To confirm the statistical significance of this, in none of 10000 random selections of 13 genes from the PIN did we observe the same level of consistency and statistical significance as for the Notch pathway genes ($P<0.0001$), indicating that reduced entropy of the Notch pathway is a key feature of the non-pluripotent state ({\bf SI , fig.S17}). It is also important to demonstrate that the interactors driving the lower entropy of Notch genes are other Notch-pathway genes. For many Notch genes (e.g. {\it NOTCH2, NOTCH3, DLL1, JAG1, PSENEN, APH1A, APH1B}) this was indeed the case, despite the fact that there were also many non-Notch pathway interactors present ({\bf SI , figs.S16,S18}).\\
To further test the added value of local network entropy, we revisited the HL60 to neutrophil time course data. Using linear regressions we identified the genes showing the most significant decreases and increases in network entropy. Ranking genes according to those showing the largest reductions in network entropy and performing a subsequent Gene Set Enrichment Analysis (GSEA), we identified JAK-STAT signalling as one of the key pathways ({\bf SI , fig.S19-S20}). The involvement of this pathway is heavily supported by previous studies \cite{Minami1996,Caldenhoven1998,Kanayasu-Toyoda1999,Coffer2000}. Attesting to the statistical significance of the JAK-STAT pathway, computing entropies after randomly permuting the gene expression profiles over the nodes in the network led to no significantly enriched biological terms (adjusted P-values $>0.05$). This is an important result because it shows that the dynamic network entropy changes inferred from the integrated PIN are indeed targeting specific signaling pathways. Finally, using non-network based approaches did not identify the JAK-STAT pathway ({\bf SI , fig.S19}).

\section*{Discussion}
Here we have taken a systems analysis view of cellular differentiation, proposing the concept that network entropy is inversely correlated with the differentiation status of a sample. By computing the network entropy of over 800 samples, encompassing cell types from many diverse cell-lineages and differentiation stages, and profiled using a variety of different microarray platforms, we have demonstrated that entropy provides a near absolute quantification measure of the undifferentiated state of any given sample.\\
In the context of normal physiology, hESCs and other pluripotent cell types were correctly predicted to exhibit the highest levels of network entropy, followed by multipotent stem cells (e.g. NSC/HSC/MSC), with terminally differentiated cells exhibiting significantly lower entropy ({\bf Fig.5}). In the context of cancer, CSCs exhibited higher levels of cellular entropy than ordinary cancer cells, although this difference appears substantially reduced in comparison to what is observed between normal stem cells and their differentiated progeny ({\bf Fig.5}). Cancer cell lines exhibited a higher entropy than primary cancers, with cancer tissue possessing higher values than normal tissue ({\bf Fig.5}).

\begin{figure}[ht]
\begin{center}
\includegraphics[scale=0.75]{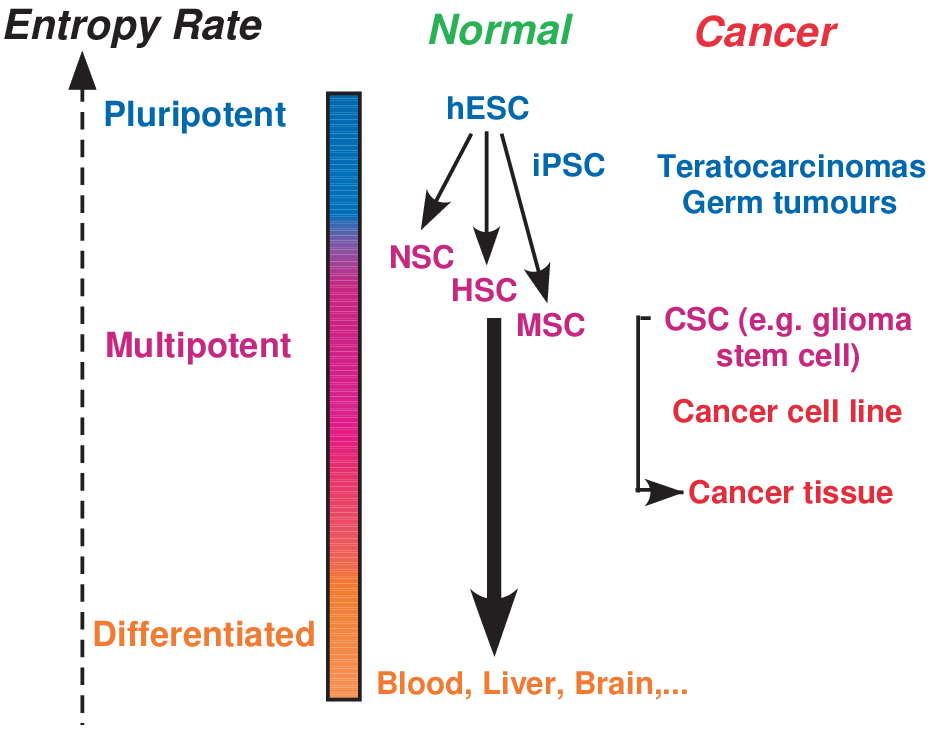}
%
%
\caption{{\bf Network entropy rates between major cell types in normal and cancer physiology:} Network entropy correlates with pluripotency and anticorrelates with the differentiation stage of cells. hESC: human embryonic stem cell, iPSC: induced pluripotent stem cell, MSC: mesenchymal stem cell, HSC: hematopoietic stem cell, NSC: neural stem cell, CSC: cancer stem cell.}
\label{fig:5}       
\end{center}
\end{figure}

All these findings are consistent with network entropy being a direct measure of the average intra-sample cellular heterogeneity, supporting the view that cellular states such as pluripotency are a statistical property of a cell population \cite{Chang2008,MacArthur2013}. Indeed, although we have not analysed genome-wide single-cell expression data, it is highly plausible that the degree of cellular heterogeneity is determined by the level of signaling promiscuity, and hence stochasticity, in single cells \cite{Chang2008,MacArthur2013}. The observation that cancer stem cells exhibit a high but marginally lower network entropy than their normal stem cell counterparts is also consistent with the view that CSCs must be characterised by oncogenic pathway dependencies, which, as shown in a previous study, lead to a lowering of network entropy \cite{West2012}. Local entropy analyses aimed at identifying the specific oncogenic pathways driving the lower entropy in CSCs could thus offer novel therapeutic opportunities \cite{West2012}.\\
It is important to stress again that network entropy provides a very general system's measure of the undifferentiated state of a sample. In this regard, we remark that reported pluripotency expression signatures \cite{Mikkelsen2008,Palmer2012}, which lack a systems-level interpretation and understanding, could only consistently discriminate pluripotent from non-pluripotent cell types, but generally failed to discriminate cell types located further down the differentiation hierarchy, irrespective of normal or cancer physiology ({\bf SI , figs.S21-S27}). Thus, the fact that network entropy provides a more refined classification of the distinct cell types across the global differentiation hierarchy, and that it did so independently of cell-proliferation indices, attests to the biological importance of this measure and of the statistical mechanical framework on which it is based.\\
Although we observed some variation in entropy rates between studies profiling the same cell-types using the same technology, it is nevertheless also important to note that these variations were in general small and that network entropy provided a relatively robust measure of the undifferentiated cellular state: for instance, hESCs always showed the highest levels of network entropy, irrespective of study or platform. This robustness stems from two key features. First, network entropy is a self-calibrating measure, as it is constructed by taking ratios of gene expression intensity values. This makes it a dimensionless quantity and fairly insensitive to the microarray or normalisation method used, unlike the scores derived from pluripotency signatures which showed significant variations between studies (see {\bf SI , fig.S28}). Second, network entropy is not affected by overfitting since it is a quantity which does not depend on feature selection. Thus, unlike pluripotency expression signatures \cite{Mikkelsen2008,Palmer2012,Mueller2011}, network entropy does not depend on tunable parameters. It follows that network entropy could provide a simple, general and robust quantitative test for assessing the pluripotency or multipotency of a cellular sample. For instance, it could be used to assess the quality of iPSCs in reprogramming experiments or even to identify mislabeled samples.\\
Since a sample's network entropy is computed from integration of its genome-wide expression profile with a protein interaction network, it is important to also comment on the robustness of the results in relation to the network, and more importantly on the number of genes that are measured. Considering the HL60 differentiation time course data set as a test case, we observed that randomly subsampling from the underlying integrated network and recomputing the entropy rates for the resulting maximally connected components, still resulted in significant decreases of the entropy rate with differentiation stage, as long as we subsample at least 40$\%$ of genes in the network ({\bf SI, fig.S29}). That the association between network entropy and differentiation stage is robust to subsampling indicates that the dynamic changes in entropy are driven by a subtle interplay between the gene expression changes and the topological properties of the nodes exhibiting these changes. We leave investigation of this and other aspects to a future study.\\

In summary, we have proposed a relatively simple, computable, systems property of a genome-wide expression profile, called network entropy, which provides an estimate of signaling promiscuity and cellular heterogeneity, and which correlates with the undifferentiated state of cells. Network entropy may thus serve as a quantitative {\it in-silico} proxy for a sample's differentiation potential in Waddington's epigenetic landscape.

\section*{Methods}

Full details of the data sets, interaction network and all statistical methods used are provided in {\bf SI}. Below, we give a brief sketch of how network entropy is calculated.

\subsection{Construction of the sample specific stochastic matrix and network entropy rate}
The sample specific stochastic matrix is estimated by integrating the gene expression profile of the sample with a comprehensive protein interaction network. Specifically, we invoke the mass action principle: let $E_i$ denote the normalised expression level of gene $i$ in a given sample. For a given neighbour $j\in N(i)$ (where $N(i)$ labels the neighbours of $i$ in the PIN), the mass-action principle states that the probability of interaction with $i$ is approximated by the product $E_{i}E_{j}$. Normalising this to ensure that $\sum_{j}p_{ij}=1$, we get
\begin{equation}
p_{ij}=\frac{E_{js}}{\sum_{k\in N(i)}E_{ks}} \qquad\forall j\in N(i)
\label{eq:stoch1}
\end{equation}
Clearly, if $j\notin N(i)$, then $p_{ij}=0$. This then defines a sample-specific stochastic matrix. From this stochastic matrix one can then construct a local network entropy for each gene $i$ in the PIN, as
\begin{equation}
S_i = -\sum_{j\in N(i)}{p_{ij}\log{p_{ij}}}
\end{equation}
which reflects the level of uncertainty or promiscuity in the local interaction probabilities around gene $i$. We note that the above expression for the local entropy is not normalised so that the maximum possible entropy depends on the degree ($k_i$) of the node $i$. In fact, $\max{S_i}=\log{k_i}$. Thus, it is convenient to also define a normalised local entropy as (see \cite{Teschendorff2010bmc}),
\begin{equation}
\tilde{S}_i = -\frac{1}{\log{k_i}}\sum_{j\in N(i)}{p_{ij}\log{p_{ij}}}
\end{equation}
We stress again that this local network entropy can be computed for each gene $i$ in each given sample. When defining a global network entropy (i.e. for the whole network) one can, in principle, consider the average of these normalised local entropies. This average however is a non-equilibrium entropy \cite{West2012}, in contrast to the global entropy rate, $S_R$, which is defined in terms of the stationary distribution, $\pi$, of the stochastic matrix $p$, i.e. through $\pi p = \pi$. Specifically, the global entropy rate, $S_R$, is defined by \cite{Latora1999,GomezGardenes2008}
\begin{equation}
S_R=\sum_i{\pi_iS_i}
\end{equation}
where $S_i$ are the unnormalised local entropies. We note that the network entropy rate is bounded between 0 and a positive maximum value that depends only on the adjacency matrix of the network \cite{Demetrius2005}. Indeed, it can be shown that the maximum possible entropy rate is attained by a stochastic matrix, $p_{ij}$, defined by $p_{ij}=A_{ij}v_j/\lambda v_i$, where $A_{ij}$ is the adjacency matrix (i.e. unweighted) of the PIN, and $v$ and $\lambda$ are the dominant eigenvector and eigenvalue of this adjacency matrix, respectively. The maximum attainable entropy rate, $M_R$, will thus depend on the specifics of the network, including total number of genes, edges and topology. Thus, to facilitate comparison between networks, the network entropy rate, $S_R$, can be scaled relative to the maximum attainable value in that given network, $\tilde{S}_R\equiv S_R/M_R$, so that $\tilde{S}_R$ is always bounded between 0 and 1. In this work, all reported entropy rates have been normalised in this way.\\
We note that computation of the entropy rate is computationally intensive as it requires estimation of the stationary distribution of a large stochastic matrix. For a connected network of size 8290 nodes, computation of a sample's entropy rate takes $\sim 10$ minutes on a Dell Precision T5400 workstation. R-scripts performing the computations are freely available on request.

\section*{Acknowledgments}
CRSB was supported by a EPSRC/BBSRC CoMPLEX PhD studentships. AET was supported by a Heller Research Fellowship. We would like to thank Roger Kramer for data collection tasks, and Alex Gutteridge for data pointers.

\section*{Author contributions}
Statistical analysis was performed by CRSB and AET. Study was conceived by AET, CRSB and JZ. AET wrote the manuscript with contributions from CRSB. DMS contributed data. SS contributed funding. MW and TE contributed ideas.

\section*{Conflict of Interest}
The authors declare that they have no conflict of interest. The funders had no role in study design, data collection and analysis, decision to publish, or preparation of the manuscript.

\section*{References}

\bibliographystyle{nature}

\end{document}